\documentclass[12pt]{iopart}

\usepackage[utf8]{inputenc}
\usepackage{graphicx}
\usepackage{mciteplus}
\usepackage{nicefrac}

\newcommand*{\tinybee}{\textit{TinyBee}}
\newcommand*{\snse}{18R-SnSe$_{2}$(CoCp$_{2}$)$_{0.19}$}
\newcommand*{\snsefull}{18R-SnSe$_{2}\lbrace\textrm{Co}(\eta{}\textrm{-C}_{5}\textrm{H}_{5})_{2}\rbrace_{0.19}$}

\begin{document}

\title[Low field extension for magnetometers (\tinybee)]{Low field extension for magnetometers (\tinybee) used for investigations on low-dimensional superconductors with $B_{\rm c1} <$ 5\,G}

\author{Manuel Presnitz}
\author{Michael Herzinger}
\author{Ernst-Wilhelm Scheidt}
\author{Wolfgang Scherer\footnote{wolfgang.scherer@physik.uni-augsburg.de}}
\address{Lehrstuhl Chemische Physik und Materialwissenschaften (CPM), Institut für Physik, Universität Augsburg, Universitätsstraße 1, 86159 Augsburg, Germany}
\author{Michael Baenitz}
\address{Max-Planck-Institut für Chemische Physik fester Stoffe, Nöthnitzer Str. 40, 01187 Dresden, Germany}
\author{Michael Marz}
\address{Physikalisches Institut, Universität Karlsruhe, Wolfgang-Gaede-Str. 1, 76131 Karlsruhe, Germany}

\date{\today}

\begin{abstract}
In this article a simple and easy to install low magnetic field extension of the SQUID magnetometer \textit{Quantum Design} MPMS-7 is described.
This has been accomplished by complementing the MPMS-7 magnet control system with a laboratory current supply for the low magnetic field region ($B \leq 200\,$G).
This hard- and software upgrade provides a significant gain in the magnetic field accuracy up to an order of magnitude compared with the standard instrument's setup and is improving the resolution to better than 0.01\,G below 40\,G.
The field control has been integrated into the \textit{Quantum Design MultiVu} software for a transparent and user-friendly operation of this extension.
The improvements achieved are especially useful, when low magnetic field strengths ($B < 1$\,G) are required at high precision.
The specific advantages of this application are illustrated by sophisticated magnetic characterisation of low-dimensional superconductors like Sc$_3$CoC$_4$ and SnSe$_{2}\lbrace\textrm{Co}(\eta{}\textrm{-C}_5\textrm{H}_5)_2\rbrace_x$.
\end{abstract}

\pacno{07.55.-w; 74.78.-w}

\maketitle

\section{Introduction}
The generation of small magnetic fields in a highly reproducible manner is one of the key prerequisites to study the magnetic properties of superconductors. Especially the precise determination of $B_{\rm c1}$ is of ultimate importance in case of the characterisation of low-dimensional, granular or organic superconductors \cite{Koncz91,Papag06,Wanka96}. Furthermore, small magnetic fields may play a crucial role in the identification and verification of the paramagnetic Meissner effect in high-temperature superconductors (\textrm{e.\,g.}\@ in Bi-2212) \cite{Svedl89,Braun92}. However, the need for reliable low magnetic fields goes beyond the mere characterisation of superconducting substances, as these also play a role in the determination of the remanent magnetisation of materials used in the construction of cryogenic instruments \cite{Meste96}.

As an application example of our hardware extension, we will show precise magnetic case studies of low-dimensional superconductors, namely the quasi one-dimensional scandium carbide Sc$_3$CoC$_4$, which was recently discovered to be superconducting \cite{Scher10,Schei11} and the metal dichalcogenide \snsefull{}.
Precise magnetisation measurements $M(B)$ at 1.8\,K and zero-field-cooling / field-cooling studies $\chi$($T$) of Sc$_3$CoC$_4$ indicate magnetic flux trapping even below 0.15\,G.
In case of the layered superconductor \snsefull{}
we demonstrate by applying low magnetic fields with minute changes (increments $\Delta B = 0.25$\,G) that the accuracy of our hardware extension is precise enough to identify strong flux pinning above $B = 0.8$\,G.\footnote{\label{Fn:SI}Although SI units were employed throughout, the
unit Gau\ss{} was used for convenience; $1\,\textrm{G}\equiv 10^{-4}\,\textrm{T}$.
We also stick to the unit emu for the magnetic moment as given by the MPMS measurement software; $1\,\textrm{emu}\equiv 10^{-3}\,\textrm{Am}^2$.
The term \emph{magnetic field} is used for the magnetic induction $B = \mu_{\rm{0}}H$.}

These precise magnetisation measurements were achieved by designing a low magnetic field extension for the commercial available SQUID-based magnetic property measurement system \textit{Quantum Design} MPMS-7. This hardware upgrade will be denoted \tinybee{} in the following.\footnote{The name \tinybee{} is intended as a pun for ``tiny $\vec B$'', \textit{i.\,e.\@} low magnetic fields.} The intention of this extension is not to compete with state-of-the-art commercial systems or completely custom tailored research instruments, but to offer a low cost upgrade for the numerous research groups already operating a similar instrument as described herein.

The SQUID-magnetometer (MPMS-7) in its standard configuration is able to provide field increments as low as 0.1\,G with a typical relative deviation of the order of 10\% for $B = 1$\,G when using the superconducting magnet coil.%
	\footnote{The presented extension uses the superconducting magnet coil and not the optional AC primary coils as described in the \textit{Quantum Design} MPMS Application Note 1014-212. Accordingly, no (potentially risky) modification of the sample tube is necessary, as described in Ref.\,\cite{Wang93}.}
Since the setup of the magnetic field control of the MPMS-7 is well-arranged, and the control software \textit{MultiVu} allows the incorporation of custom-tailored modules via the \textit{external device control} (EDC) interface, it was possible to extend the magnetic field control with a more accurate current source and to achieve a seamless integration into the normal \textit{MultiVu} measurement interface. The extension described herein consists of a hardware modification (chapter 2) and a software control module (chapter 3). The discussion of the accuracy and reproducibility of the magnetic field values is completed by a comparison of our test case magnetisation measurements with alternative commercial systems (chapter 4). Finally, we discuss our new results on the flux trapping behaviour in the low-dimensional superconductors Sc$_3$CoC$_4$ and \snsefull{} using the \tinybee{} option (chapter 5).

\section{Hardware Extension}
The original current source (Kepco model JQE 6--45) is used to apply electric currents up to 45\,A to achieve magnetic fields as large as 70\,kG. Unfortunately only two current ranges are implemented into the magnetic control system; the current is sensed across a common shunt resistor of 0.1\,$\Omega$ and 0.01\,$\Omega$  for the low and high magnetic field region, respectively. The voltage drop is digitalized with 16\,bit resolution, \textit{i.\,e.\@} the lowest significant bit (LSB) corresponds to a current change of $\Delta I_{\rm LSB}\geq 55\,\mu$A and therefore to a change of the magnetic field of $\Delta B_{\rm LSB}\geq 0.11\,$G. This lower bound results if one assumes that the maximum input of the ADC is 0.36\,V, which is the voltage drop at the upper limit of 7500\,G of the low current mode.
Hence, it is evident that this setup does not meet the requirements of precise magnetic studies at low magnetic fields smaller than 1\,G.

We have therefore complemented this standard device by a laboratory current source ($I\le 100\,\textrm{mA}$), and an automatic switching assembly between high field ($B> 200\,\textrm{G}$) and low field operation ($B\le 200\,\textrm{G}$).
Our hardware of choice is a discontinued \textit{Keithley} Model 224, but other current sources, \textit{e.\,g.\@} the current \textit{Keithley} Model 6220, are also suitable for this purpose. The used device is programmable via an IEEE-488 interface and provides also four digital output channels, which are used for operational as well
as control purposes as described below.

\begin{figure}[!h]
	\centering
	\includegraphics[width=0.75\textwidth]{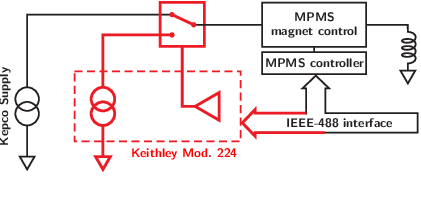}
	\caption{Block diagram of the hardware modification used as extension of the magnetic field control of the MPMS-7. Added components are depicted in red and are emphasized (colour online). The changeover switch is realized through four relays.}
	\label{Abb:Blockschaltbild}
\end{figure}

Only minor hardware modifications are necessary as outlined in the block diagram (Fig.\,\ref{Abb:Blockschaltbild}). In order to provide an automatic switching between the current sources we patched a high current relay into each of both current supply lines (\textit{approx.} AWG 9) from the Kepco power supply to the magnet interface in the MPMS-7 cabinet.\footnote{As an economic alternative for the required high current switches so-called plug-in relays, normally used in automotive engineering, were employed.} In addition, both terminals of the magnet interface are wired (AWG 21) over two DIL relays, located on a perforated board, to the \textit{Keithley} current source. Mandatory fuses (63\,mA, \textit{flink} characteristic) are looped in to cut off the low current part from the high current supply as fast and as early as possible in case of a short circuit. Both relay pairs together act as a changeover switch (like it is depicted in Fig.\,\ref{Abb:Blockschaltbild}). They are excited via two driving stages which are controlled by two of the digital output ports of the \textit{Keithley} current source and hence are accessible over the IEEE bus.
At least two independent output ports are necessary as the relays controlling the positive and negative lines can be switched simultaneously, but the high current path should be interrupted before the low current part gets attached.
The remaining two output ports of this \textit{Keithley} device are used to connect indicator lights which are mounted on the front panel of the MPMS-7 device.

A detailed schematic diagram can be obtained from the authors.%
\footnote{\label{Bezug}%
	Please contact us directly via electronic mail at manuel.presnitz@physik.uni-augsburg.de.
	Be aware, that the authors take no liability or responsibility whatsoever for any kind of injuries, damages or malfunctions that may occur during the installation or use of the modification described herein.}

\section{Software Extension}
The control module for our low field extension was written in the Delphi programming language (Borland Delphi 5) as required by the \textit{external device control} (EDC) interface of the \emph{MultiVu} software of the MPMS-7. All input commands are controlled via a parameter string within the EDC function call, which again is included in the measurement sequence file. Syntax and available commands are specified in detail in the documentation.

In the standard MPMS-7 setup, the \textit{No Overshoot}
mode is used for applying low fields. Hence, we developed a monotonic ramp function which sweeps to the field current for the desired magnetic field in roughly 100\, sec. This time period is only weakly dependent on the magnitude of the change. Furthermore it is guaranteed that the set point of the magnetic field is never exceeded. Implementation of additional delays for the relay switching time, the current settling or the change of the persistent switch heater state was also necessary. To determine the correct timings we analysed the IEEE bus activity during a magnet field change using the original MPMS-7 magnet control and also monitored directly the state of the controller's output lines.
We further note, that the \tinybee{} software module has been optimized for low activity on the IEEE bus. This is achieved by emulation of the smallest current steps of which the \textit{Keithley} current supply is capable, and by sending a new value to the device only in those cases where this procedure would yield a current change.

Worth noting is also the additional implementation of two types of control loops: the \textit{repeat} and the \textit{do} loop. Together with the automatic range selection, \textit{i.\,e.\@} all magnetic field values can be set via the \tinybee{} EDC module while values $B>200$\,G are passed directly to the MPMS-7 controller, this offers the possibility to run field-dependent measurements rather comfortably over a wide range of magnetic field strengths.

Source code%
\footnote{A library for parsing regular expressions from Andrey V. Sorokin was used and is copyrighted by him: A. V. Sorokin, \emph{TRegExpr class library} (2004), http:$\slash\!\slash$RegExpStudio.com}
and documentation are available upon request from the authors; see footnote \ensuremath{\ref{Bezug}} on page \pageref{Bezug}.

\section{Performance of the system}

\subsection{Calibration of the fluxgate probe}
At room temperature a 10 Gauss fluxgate probe (available from \textit{Quantum Design}) was used to determine the actual magnetic field strength $B_{\rm act}$ at the sample position inside the sample tube.
First, we verified the performance of the probe with a self wound copper coil (solenoid geometry: diameter \textit{approx.}\@ 15\,mm and 1300 windings  per 154.3\,mm of enamel-coated copper wire, \o{} = 0.10\,mm), with an estimated error of 1\,\% for the generated magnetic field. Within this error margin the reading of the fluxgate probe is correct.
We then checked the performance of the \tinybee{} setup: The dependence of the actual values from the set values turned out to be highly linear, \textit{i.\,e.\@} the deviation of each data point from linearity lies within the verified accuracy of the fluxgate probe over the whole measuring range ($-10$ to $+10$\,G).
However, we observe a systematic deviation of \textit{approx.}\@ 4\,\% of the actual field $B_{\rm act}$ from the applied magnetic field $B_{\rm app}$ within this range. This error can be reproduced using the \tinybee{} setup as well as the built-in MPMS-7 magnet control. We therefore suppose a slightly field dependent variation at low fields of the proportionality factor between current and magnetic field of the superconducting field generating coil. This assumption is based on the fact that the proportionality factor has to be calibrated at fields $B_{\rm app} > 10$\,kG, in order that ferromagnetic impurities in the employed palladium reference sample (provided by \textit{Quantum Design}) can be neglected. \cite{QD1041-001-pub}

\subsection{Comparison of the standard MPMS-7 setup with the \protect\tinybee{} option using a type-I superconductor (Pb)}
\label{Chapter:Zerofield}
We also checked the performance of our \tinybee{} setup with a spherical lead ball (Strem Chemicals, purity 6N, $m = 198.32$\,mg, $d\approx 3.2$\,mm). The critical temperature of this type-I superconductor is $T_\textrm{c} = 7.2$\,K.
Below $T_\textrm{c}$ a high voltage is induced in the pick up coils, indicating a strong (negative) magnetisation. In this case the fit routine, which assumes an ideal dipole sample, can determine the magnetic moment at high precision.
For an ideal type-I superconductor the presence of vortex dynamics or flux trapping can be safely neglected.
Therefore, in the superconducting state, the volume susceptibility $\chi_V$ equals $-1$. Hence, the relation of the actual magnetic field $B_{\rm act}$ and the measured magnetic moment $M_{\rm P}$ per sample reduces to
\begin{equation}
	\frac{B_{\rm act}}{\mu_0} = \left( N + \frac{1}{\chi_V} \right) \frac{\rho}{m} M_{\rm P} = (N-1) \frac{\rho}{m} M_{\rm P}.
	\label{Gl:MagMoment}
\end{equation}
Here $N$ is the demagnetisation factor ($N=\nicefrac{1}{3}$ for spherical samples), $\rho$ the density ($\rho = \rho_{\rm Pb} = 11.34$\,g/cm$^3$), and $m$ the mass of the sample.

The following procedure was used to estimate the reproducibility of the applied magnetic field $B_{\rm app}$:
At the beginning the superconducting magnet was quenched to eliminate the magnetic flux trapped in the coil. Then, after applying a counter field to compensate $B_0$, \textit{i.\,e.\@} the earth's magnetic field together with possible remaining flux in $z$ direction, $B_0 = 0.66$\,G (measured at 300\,K with the fluxgate probe), the sample was cooled down to 2\,K in zero magnetic field.

Subsequently,
(\textit{i}) a field value of $B_{\rm eff} = B_{\rm app} + B_0 = 1$\,G was achieved;
(\textit{ii}) after stabilization for 2~minutes, the magnetic moment of the sample was recorded;
(\textit{iii}) the magnetic field was reduced to $B_{\rm eff} = 0$\,G and after another delay of 2~minutes $B_{\rm eff} = 1$\,G was established again.
By repeating these three steps, ten data points were recorded, once with our new \tinybee{} setup and once again with the built-in MPMS-7 field control system.\footnote{For the MPMS-7 setup the \textit{No Overshoot} mode was used with \textit{Hi Res} option enabled. $B_{\rm app} = 1\,\textrm{G}-B_{0} = 0.34$\,G had to be rounded to 0.3\,G.} For the latter setup the resulting values are noted below in square brackets.
The mean value of the measured magnetic moment $M_{\rm P}$ is $\overline M_{\rm P} = -2.138 \,[-2.209]$\,memu, with a standard deviation of $1.67\,[14.7]\,\mu$emu (for a remark about the unit emu see footnote \ref{Fn:SI} on page \pageref{Fn:SI}). According to Eq.\,(\ref{Gl:MagMoment}) these values correspond to the actual magnetic field of $B_{\rm act} = 1.024$\,[1.058]\,G with a standard deviation 0.8\,[7.0]\,mG. The improvement of the reproducibility  becomes clearer by calculating the range of the respective data set, which gets reduced by one order of magnitude from 21\,mG for the built-in MPMS-7 field control system to 2.1\,mG for the \tinybee{} setup.

In addition, we repeated this procedure for different magnetic fields.
For better comparison we have chosen magnetic field values $B_{\rm app}$ which can be selected identically with both current supply setups, as the MPMS-7 setup is fixed to values that are multiples of 0.1\,G. Together with $B_0$, in this case $B_{0} = 0.57$\,G,%
\setcounter{footnote}{1}\footnote{$B_0$ includes not only the earth's magnetic field, but also remaining flux in $z$ direction and hence can vary about \textit{approx.\@} 0.1\,G.}
the effective magnetic field values $B_{\rm eff} = B_{\rm app} + B_{0}$ are obtained and are listed in Tab.\,\ref{Tab:Tiny-vs-Kepco-Std}. $B_{\rm act}$ values follow again from Eq.\,(\ref{Gl:MagMoment}), \textit{i.\,e.\@} $\chi_{\rm V} = -1$ is assumed; the mean value for the ten data points is denoted $\overline{B}_{\rm act}$. The table is completed by the relative error $\overline{B}_{\rm act}/B_{\rm eff}-1$ and the relative standard deviation (RSD). The latter is the standard deviation with respect to $\left|\overline{B}_{\rm act}\right|$ and is therefore a measure of the dispersion of the actual field values reached in different cycles for the same $B_{\rm eff}$ value.

\begin{table}
\centering
\begin{tabular}{cccccccccc}
	\hline\hline
	& & & \multicolumn{3}{c}{\tinybee} & & \multicolumn{3}{c}{\textit{MPMS-7}}\\
	$B_{\rm app}$ & $B_{\rm eff}$
	& \quad{} &
	$\overline{B}_{\rm act}$ & rel. error & RSD
	& \quad{} &
	$\overline{B}_{\rm act}$ & rel. error & RSD\\
	(G) & (G)
	& &
	(G) & (\%) & (\%)
	& &
	(G) & (\%) & (\%)\\
	\hline
	$-0.4$ & 0.17	&	&	0.161	&	$-5.3$	& 0.52 &	&   $-0.067$	&	$-139$ & 119\\
	$-0.2$ & 0.37	&	&	0.363	&	$-1.9$	& 0.10 &	&	0.115	&	$-69$ & 4.4 \\
	0.0 & 0.57	&	&	0.566	&	$-0.7$	& 0.10 &	&	0.326	&	$-43$ & 1.6\\
	0.2 & 0.77	&	&	0.768	&	$-0.3$	& 0.01 &	&	0.872	&	13 & 11\\
	0.4 & 0.97	&	&	0.971	&	0.1	& 0.03 &	&	1.02	&	5.2 & 5.4 \\
	0.9 & 1.47	&	&	1.48	&	0.7	& 0.03 &	&	1.57	&	6.8 & 0.21\\
	1.4 & 1.97	&	&	1.98	&	0.5	& 0.02 & & 1.99	&	1.0 & 3.4\\
	4.4 & 4.97	&	&	5.02	&	1.0	& 0.04 & &	5.05	&	1.6 & 0.13\\
	\hline\hline
\end{tabular}
\caption{Deviation of the determined $\overline{B}_{\rm act}$ values from the applied effective field values $B_{\rm eff} = B_{\rm app}+B_0$ for different magnetic fields employing both current supply units. Values given for $\overline{B}_{\rm act}$ are calculated according to Eq.\,(\ref{Gl:MagMoment}) and represent mean values of 10 data points. The relative errors follow from $\frac{\overline{B}_{\rm act}}{B_{\rm eff}}-1$ and the relative standard deviation (RSD) is the standard deviation with respect to the actual field $\left|\overline{B}_{\rm act}\right|$.
}
\label{Tab:Tiny-vs-Kepco-Std}
\end{table}

\begin{figure}
	\centering
	\includegraphics[scale=0.98]{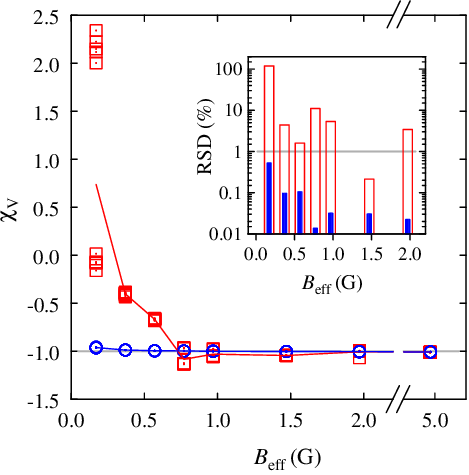}
	\caption{Dependency of the volume susceptibility $\chi_{\rm V}$ \textit{vs.\@} the effective magnetic field $B_{\rm eff}$. To test the reproducibility of the applied magnetic field at different field values, 10 data points were recorded each for a spherical lead ball at 2\,K (see text). Data points recorded with the \tinybee{} setup and the built-in MPMS-7 field control device are depicted by circles and squares, respectively. The solid lines connect the mean values of the respective data points. In the inset the relative standard deviation (RSD), \textit{i.\,e.\@} the standard deviation with respect to $\left|B_{\rm act}\right|$, is shown in a semi-logarithmic plot with full and empty bars for the \tinybee{} setup and the built-in MPMS-7 field control, respectively.}
	\label{Abb:Tiny-vs-Kepco-Std}
\end{figure}

On the contrary, Fig.\,\ref{Abb:Tiny-vs-Kepco-Std} illustrates the error arising in a practical application. Here the volume susceptibility $\chi_{\rm V}$ is plotted \textit{vs.\@} $B_{\rm eff}$, where $\chi_{\rm V}$ is calculated under the commonly used assumption that the actual magnetic field $B_{\rm act}$ is identical with the applied effective field $B_{\rm eff}$,
\begin{equation}
	\chi_{\rm V} = \left(\frac{m}{\rho}\frac{B_{\rm eff}}{\mu_0 M_{\rm P}} - N\right)^{-1}.
	\label{Gl:ChiV}
\end{equation}

For fields smaller than 1\,G, the absolute deviation from the ideal value $\chi_{\rm V} = -1$ increases dramatically for the built-in MPMS-7 field control device. Eventually for the smallest field ($B_{\rm eff} = 0.17$\,G) we even observed a change of sign for some of the data points; this is caused by the overcompensation of $B_0$. Therefore, the built-in MPMS-7 field control device is virtually unsuitable for precise measurements below 1\,G. By contrast the alternative \tinybee{} setup exhibits an excellent  accuracy as witnessed by small absolute deviations (see Tab.\,\ref{Tab:Tiny-vs-Kepco-Std}) in combination with a good field reproducibility. Even though the RSD values increase with decreasing field values, they stay well below 1\% (inset of Fig.\,\ref{Abb:Tiny-vs-Kepco-Std}).

\begin{figure}
	\centering
	\includegraphics[scale=0.98]{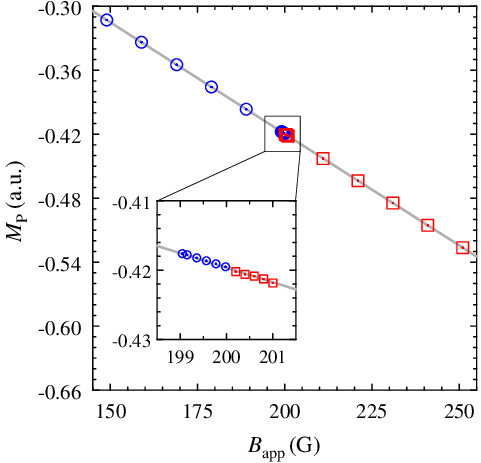}
	\caption{The automatic switching between both current sources at 200\,G is illustrated by the dependency of the magnetic moment $M_{\rm P}$ of the lead ball \textit{vs.\@} the applied magnetic field $B_{\rm app}$. The smoothness of the switching is better recognizable in the inset, which zooms into the region around 200\,G. Data points for the \tinybee{} and MPMS-7 setup are depicted with circles and squares, respectively. The line represents a linear fit through all data points.}
	\label{Abb:Tiny-with-Kepco}
\end{figure}

In Fig.\,\ref{Abb:Tiny-with-Kepco} the measured magnetic moment per sample $M_{\rm P}$ (using the lead ball described above) \textit{vs.\@} the applied magnetic field $B_{\rm app}$ is shown. This demonstrates that the values of $M_{\rm P}$ display a smooth and continuous transition when the magnetic field exceeds the maximum value feasible by the \tinybee{} setup. In such a case, the software automatically switches to the built-in MPMS-7 current supply for fields greater than 200\,G. Hence, a measurement sequence can be carried out easily over wide magnetic field ranges employing our \tinybee{} control software without any manual range selection.

\clearpage
\subsection{Comparison of the \protect\tinybee{} setup with different commercial magnetometers}
The superconducting compound Sc$_3$CoC$_4$, described in more detail in the next chapter, had been selected as a benchmark system for a comparison of different commercially available magnetometers. A spherical sample ($m = 152.67$\,mg, $d\approx 4.5$\,mm) was cooled in zero magnetic field down to $T = 1.8$\,K to record an initial magnetisation curve. The field increment was chosen to be very small ($\Delta B \leq 0.15$\,G) to demonstrate the high accuracy of the \tinybee{} setup.

\begin{figure}[b]
	\centering
	\includegraphics{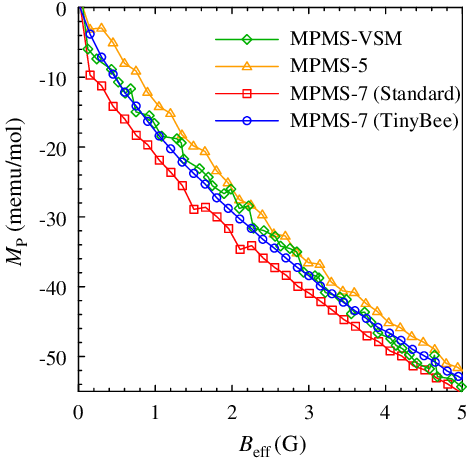}
	\caption{Field dependent magnetisation (shown in the range between 0 and 5\,G) recorded with different commercial magnetometers. All measurements were performed on the quasi-one-dimensional superconductor Sc$_3$CoC$_4$ (with \emph{approx.\@} spherical shape) at $T = 1.8$\,K.}
	\label{Abb:Vergleich_Geraete}
\end{figure}

In addition to the MPMS-7 SQUID-magnetometer (with the low field extension \tinybee{} setup and with the built in power supply) two alternative commercial systems were used for comparison, a MPMS-5 and the state-of-the-art SQUID-VSM dc-magnetometer, both from \textit{Quantum Design}.
The resulting field dependent magnetisation measurements  are shown in Fig.\,\ref{Abb:Vergleich_Geraete} for $B < 5$\,G. All data points were collected stepwise in the \textit{No Overshoot} mode at constant magnetic fields, with exception of the data points extracted from the SQUID-VSM system. Here, the measurements were performed in the \textit{Continuous Field Sweep} mode, because of the very fast VSM technique (one magnetisation point takes 1\,sec). For high accuracies we selected 0.1\,G/sec as the field sweep rate. As a result, the different data sets are in very good qualitative agreement. In regarding the smoothness and the uniformity of the $M$($B$) data points the \tinybee{} curve clearly outperforms all other measurements.

\section{Experimental section}

\subsection{Missing ideal diamagnetism in the quasi-one-dimensional Sc$_3$CoC$_4$}
\label{Chapter:Sc3CoC4}

In the following we present selected case studies to demonstrate the performance of the \tinybee{} low field extension. We first focus on the rare-earth transition metal carbide Sc$_3$CoC$_4$ which represent a recent example of a low-dimensional superconductor ($T_{\rm{c}}^{\rm{onset}} = 4.5$\,K) \cite{Scher10}. Sc$_3$CoC$_4$ displays quasi-one-dimensional [CoC$_4$]$_{\infty}$ ribbons at room temperature. \cite{Rohrm07} Below 72\,K a structural phase transition occurs causing a zigzag-type chain deformation, due to the alternating displacement of the Co atoms along the [CoC$_4$]$_{\infty}$ chains. Such a zigzag-type pattern appears to be a typical structural motif displayed by quasi-one dimensional superconductors and also Sc$_3$CoC$_4$ was identified as one of the few model system showing the rare phenomenon of quasi-one-dimensional superconductivity. \cite{Schei11, Scher10c, Baeni11} Accordingly, Sc$_3$CoC$_4$ reveals all salient properties connected with a low-dimensional superconducting behaviour: (\textit{i}.) existence of an irreversible field in the magnetic hysteresis loop ($B_{\rm{irr}} = 2800$\,G), (\textit{ii}.) a broad shape of the specific heat anomaly, (\textit{iii}.) and the upturn of the $B_{\rm{c2}}$-curve at temperatures near $T_{\rm{c}}$. \cite{Schei11, Klemm85} Furthermore, the small lower magnetic field of $B_{\rm{c1}} = (4.4\pm 0.2)$\,G, due to the large London penetration depth of about $\lambda_{\rm{L}} = 9750$\,\AA{} is also another indicator for low dimensionality. \cite{Schei11}

\begin{figure}[b]
	\centering
	\includegraphics{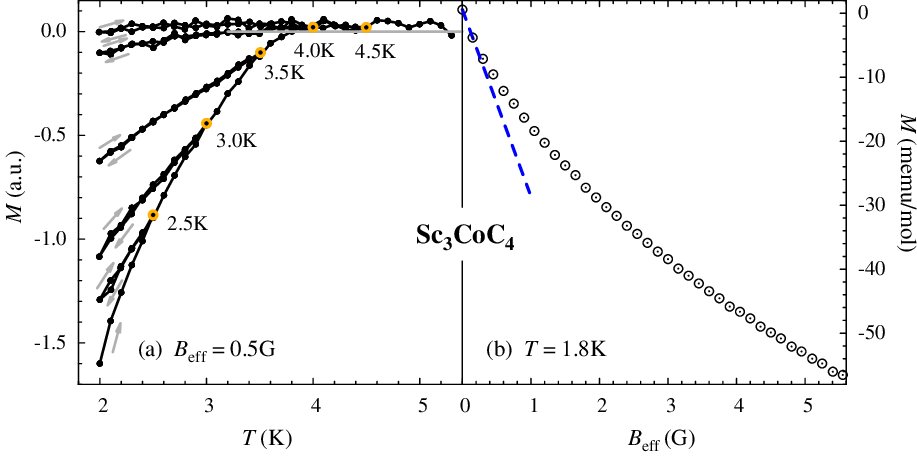}
	\caption{a) Temperature dependence of the magnetic moment of a Sc$_3$CoC$_4$ sphere in an external magnetic field $B_{\rm{eff}} = 0.5$\,G. The open circles mark the reversal points in temperature of the applied heating and cooling sequences (illustrated by arrows). b) The initial magnetisation curve $M$($B$) up to 5.5\,G of Sc$_3$CoC$_4$. The blue dashed line is a guide for the eyes through the two lowest datapoints.}
	\label{Abb:ZFC_fc_0_5G}
\end{figure}

 Fig.\,\ref{Abb:ZFC_fc_0_5G}a shows the temperature dependence of the magnetic moment $M$ of the Sc$_3$CoC$_4$ sphere in the presence of an external magnetic field of $B_{\rm{eff}} = 0.5$\,G using the \tinybee{} setup. In this case study, the earth's magnetic field and the remaining flux was compensated using the \tinybee{} setup as described in section \ref{Chapter:Zerofield} with an accuracy of $\Delta B = \pm 0.005$\,G. In the next step the sample was cooled down to 2\,K in zero field ($B_{\rm{eff}} = 0$\,G). After applying an effective magnetic field of 0.5\,G, the magnetisation was measured during the warm up sequence from 2\,K to 2.5\,K. To check the reversibility of the temperature dependent magnetisation the sample was cooled again and the magnetisation was recorded down to 2\,K. This heating and cooling sequence was repeated four times employing turnaround points at 3\,K, 3.5\,K, 4\,K and 4.5\,K, respectively. After each cycle a significant increase of the flux trapping was observed, until in the last field-cooled sequence starting from 4.5\,K the diamagnetic response vanished. In particular this experiment clearly demonstrates that magnetic flux penetrates the sample in terms of quantized vortices at magnetic fields just below the lower magnetic field of $B_{\rm{c1}} = (4.4\pm 0.2)$\,G as reported in Ref. \cite{Schei11}.

To verify this unexpected behaviour, we also performed a magnetisation measurement $M$($B$) at very low magnetic fields, where the \tinybee{} setup can make use of its inherent advantages due to its high field resolution of 0.005\,G to compensate the earth's magnetic field very precisely. Indeed, after zero-field cooling we received the first magnetic response in a field below $|B_{\rm{eff}}| < $\,0.005\,G.
In Fig.\,\ref{Abb:ZFC_fc_0_5G}b  the magnetisation \textit{vs.\@} magnetic field is plotted in the range between 0 and 5.5\,G using minute increments of $\Delta B = 0.15$\,G. The $M(B)$ dependency reveals no linear region even at lowest magnetic fields. The low noise level of this measurement is a direct result of the accurate magnetic field adjustment with \tinybee{} (cf. Fig.\,\ref{Abb:Vergleich_Geraete}). The dashed line including the datapoints for the two lowest measured fields is employed as a guide for the eyes. Even below 0.15\,G the magnetisation data deviates from the linearity, indicating flux penetration below 0.15\,G. This behaviour strongly suggests the absence of ideal diamagnetism in Sc$_3$CoC$_4$. Furthermore, the temperature dependent magnetisation measurement in Fig.\,\ref{Abb:ZFC_fc_0_5G}a proves that the magnetic flux penetrates the sample in terms of quantized flux. \cite{Schei88} In the following superconductors with such a behaviour are called \textit{early pinned superconductors}.

At this stage we should note, that for the high-temperature superconducting cuprate RuSr$_2$GdCu$_2$O$_8$ the presence of a granular structure has been proposed as origin of its missing ideal diamagnetic behaviour. The reason for this is that the pinning of the quantised vortices at low magnetic fields is mainly due to the intergrain area (area between the grains). \cite{Papag06} Therefore the missing ideal diamagnetism does not necessary hint for a lack of Meissner state.
In the case of Sc$_3$CoC$_4$, however, we cannot exclude granularity as the reason for missing ideal diamagnetism. However, a more likely explanation will be the pinning associated with the wide spacial separation in between the superconducting chains due to the embedding of the [CoC$_4$]$_{\infty}$ ribbons in the scandium matrix.

\subsection{Early flux pinning in the two-dimensional superconductor \protect\snsefull{}}

Another promising candidate for early pinning behaviour is the hybrid dichalcogenide SnSe$_2$ intercalated with 33\% cobaltocene ($\textrm{Co}(\eta{}\textrm{-C}_5\textrm{H}_5)_2$, abbreviated as CoCp$_{2}$ in the following) \cite{Forms90}.
This layered superconductor exhibits a $T_{\rm{C}} \cong 6.1$\,K and a lower critical magnetic field of $B_{\rm{c1}} \cong 10$\,G at 4.2\,K. In the following we present magnetic measurements of the layered superconductor \snse{} in the superconducting state.

\begin{figure}[b]
	\centering
	\includegraphics{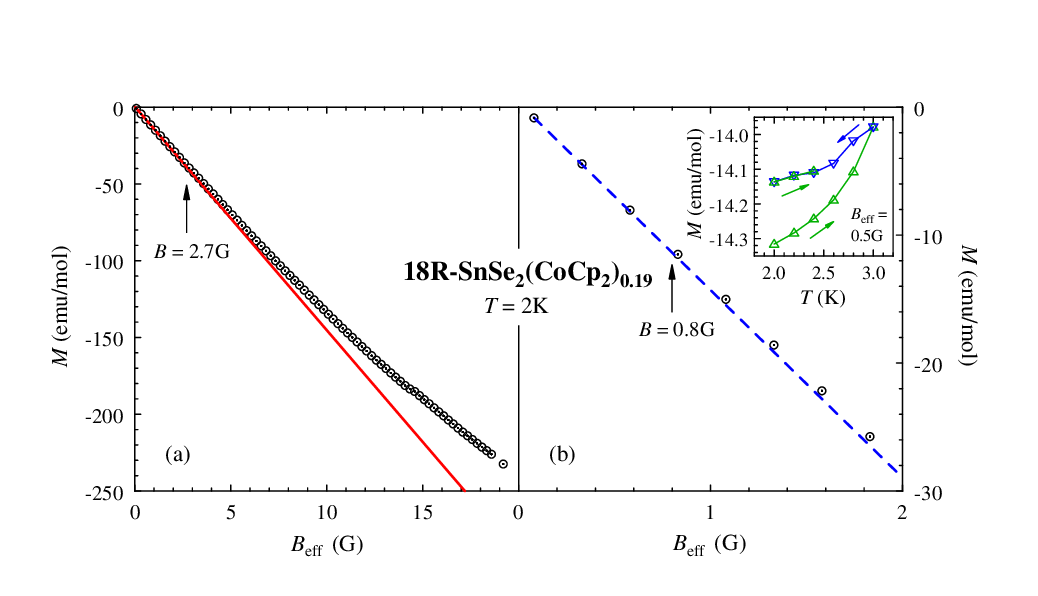}
	\caption{The initial magnetisation curves $M$($B$) up to 20\,G (a) and 2\,G (b) of \snse{}. The red line is a linear fit to the low field region, $B_{\rm eff} < 1$\,G (a) and the blue dashed line is a linear fit which only includes datapoints at even lower fields, $B_{\rm eff} < 0.5$\,G (b). The insert shows the temperature dependence of the magnetic moment of \snse{} in an external magnetic field $B_{\rm{eff}} = 0.5$\,G.}
	\label{Abb:MAG_Hc1}
\end{figure}

In Fig.\,\ref{Abb:MAG_Hc1}a the magnetisation \textit{vs.\@} magnetic field is plotted in the range between 0 and 20\,G with increments of $\Delta B = 0.25$\,G. Again we used the \tinybee{} setup to reach the highest possible accuracy. The linear fit below 1\,G deviates from the magnetisation data above 2.7\,G indicating the onset of flux penetration in connection with a small lower critical magnetic field value of $B_{\rm{c1}} \cong 2.7$\,G. The extraordinary field resolution of the \tinybee{} approach allows detailed analysis of the data below 2\,G (see Fig.\,\ref{Abb:MAG_Hc1}b). Here the dashed blue line is a linear fit which only includes datapoints at lower fields ($B_{\rm eff} < 0.5$\,G). This fit already deviates from the magnetisation curve below 0.8\,G -- a value which is significantly smaller than the estimated $B_{\rm{c1}} = 2.7$\,G. This again hints for a early pinning  behaviour which is also supported by the temperature dependent magnetisation measurement (insert of Fig.\,\ref{Abb:MAG_Hc1}b). The zero-field-cooled/field-cooled cycles, as described in chapter \ref{Chapter:Sc3CoC4}, clearly demonstrate that the  magnetic flux penetrates in terms of quantized vortexes. The origin for such early flux pinning may be intrinsic, due to the non-superconducting structural voids between the dichalcogenide layers, or extrinsic due to geometric or granular effects \cite{Papag06}.

In the following we investigated whether early flux pinning in \snse{} might be misinterpreted in terms of the presence of a paramagnetic Meissner effect \cite{Braun92} (see Supporting Information). Accordingly, magnetisation measurements were carried out in a MPMS-7 magnetometer in its standard configuration, where the sample was moved in a small field gradient of 0.05\,G/cm at 1\,G. In this case, the magnetic moment of a zero-field-cooled hard typ-II superconductor may produce an artefact, if the suppression of the earth's magnetic field to  $B_{\rm{eff}} = 0$\,G is not precisely zero. \cite{Blunt91} Especially, for an early pinned superconductor flux will be trapped during the ``zero''-field-cooling process and after moving the sample through the magnetic field gradient a paramagnetic signal may be observed. Indeed, 18R-SnSe$_{1.99}(\textrm{CoCp}_{2})_{0.1}$ displays paramagnetic behaviour after zero-field- and field-cooling when the commercial MPMS-7 setup is employed (see Supporting Information). We therefore repeated our zero-field-/field-cooled measurements after suppressing the earth's magnetic field with the \tinybee{} setup as discussed above and this time no paramagnetic response was found. Subsequent magnetic studies in a static magnetic field with a homemade SQUID and a normal conducting solenoid \cite{Vande91, Riedl94} also ruled out any presence of a paramagnetic Meissner-effect in this dichalcogenide.

This clearly affirms that samples characterized by early flux pinning demand an accurately determined efficient magnetic field, in order to avoid effects from trapped flux. We therefore recommend the usage of our \tinybee{} approach especially in cases where a paramagnetic Meissner effect is observed using the standard setup of MPMS SQUID-magnetometers. \cite{Chou93, Okram97, Soezer04, Kurod09}

\section{Concluding remarks}
Due to the complementation of the standard MPMS-7 magnetic field control with an external current supply for low fields values ($B\le 200$\,G) it is possible to increase the accuracy of applied magnetic fields by one order of magnitude and to achieve a good reproducibility. The necessary efforts are comparable small and include no critical tasks, as the hardware extension preserves the probe assembly of the standard MPMS-7 setup (in contrast to a previous reported construction \cite{Wang93}). Only modifications of the supply lines for the magnetic field coil are needed.

At the time of publication, the \tinybee{} extension has already been used and tested successfully in routine tasks for more than 24 months. With this setup it is now possible to compensate the earth's magnetic field and other residual fields very precisely via application of an opposing field. Thus  \textit{zero field cooling} measurements, which are important for investigations on superconductors or spin glass systems, become significantly more precise and facilitated.

Moreover, for the correct determination of the magnetic field where the flux penetrates the sample and of the lower critical field $B_{\rm c1}$ the possibility of applying small magnetic field with small increments together with a high accuracy and reproducibility is vital. It has been demonstrated for two examples that for samples with early flux pinning it is highly necessary to control low external magnetic fields very accurately in order to avoid misleading flux trapping effects. Therefore the \tinybee{} setup may prove in future as a very useful tool for such studies.

\section*{Acknowledgements}
The authors thank Prof. Dr. H. von Löhneysen (KIT) providing us his lab facilities to perform magnetisation measurements in a static field and Prof. Dr. A. Loidl (EKM) for experimental time at a MPMS-5 device. Furthermore, we acknowledge Dr. F. Mayr and V. Herz for fruitful discussions and W. Tratz for readily help in engineering tasks.

\section*{References}

\clearpage
\section*{Supporting Material}
\subsection*{Paramagnetic Meissner Effect (PME)}

\begin{figure}[!h]
	\centering
	\includegraphics[scale=0.75]{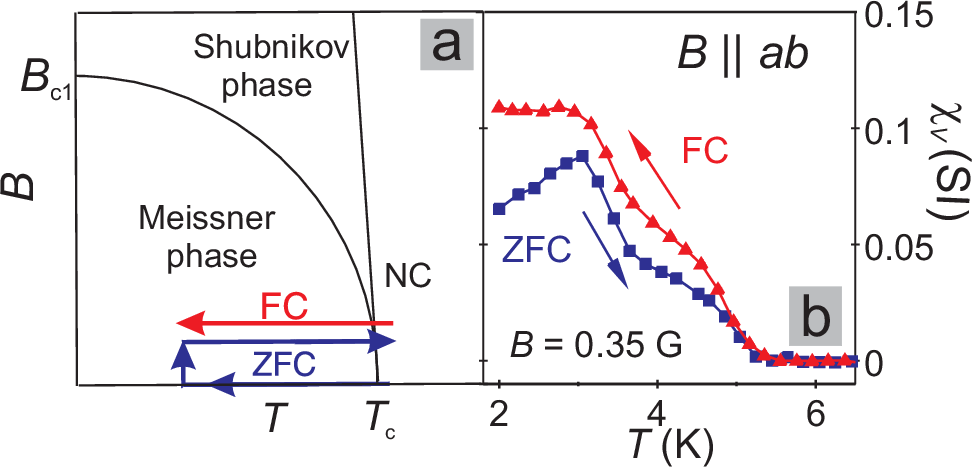}
	\caption{(a) Schematic sketch of the ZFC and FC sequences in a $B$-$T$ phase diagram; NC denotes the normal conducting state; (b) Paramagnetic response in the ZFC and FC sequences with $B$ parallel to the $ab$-planes of 18R-SnSe$_{2}(\textrm{CoCp}_2)_{0.1}$.}
	\label{Abb:PME}
\end{figure}

In the following we provide a short introduction of the PME. The best way to identify the PME is to perform magnetic field studies employing zero-field-cooled (ZFC) and field-cooled (FC) measurements. The sequences of the ZFC and FC procedures are illustrated in a schematic $B$-$T$ phase diagram (Fig.\,\ref{Abb:PME}a). ZFC sequences in low magnetic fields usually result in a complete shielding effect at low temperatures ($\chi_{V}\cong-1$), whereas the FC curves exhibit paramagnetic behavior, which decreases with increasing field due to an enhancement of the diamagnetic contribution.%
\footnote[6]{W.~Braunisch, N.~Knauf, V.~Kataev, S.~Neuhausen, A.~Gr\"{u}tz, A.~Kock, B.~Roden, D.~Khomskii, and D.~Wohlleben. Paramagnetic {Meissner} effect in {Bi} high-temperature   superconductors. {\em Phys. Rev. Lett.}, 68(12):1908--1911, 1992.}

In Fig.\,\ref{Abb:PME}b ZFC and FC dc-susceptibility curves of  18R-SnSe$_{1.99}(\textrm{CoCp}_2)_{0.1}$ in a small external magnetic field of $B\,=\,0.35$\,G are displayed. Below the superconducting transition ($T_\mathrm{c}$\,=\,6\,K) paramagnetic behavior is observed in this case for both pathways, ZFC and FC. The magnetisation measurements were carried out in a MPMS-7 magnetometer in its standard configuration, were the sample was moved in a small field
gradient of 0.05\,G/cm at 1\,G. In this case, the magnetic moment of a zero-field-cooled hard typ-II superconductor may produce an artefact,%
\footnote[7]{T.~Kuroda, T.~Nakane, and H.~Kumakura. Effects of doping with nanoscale {Co$_3$O$_4$} particles on the superconducting properties of powder-in-tube processed {MgB$_2$} tapes. {\em Physica C}, 469(1):9--14, 2009.}
when the suppression of the earth's magnetic field together with possible remaining flux in $z$ direction, $B_0$, to $B_{\rm{eff}} = B_{\rm{app}} + B_0 = 0$\,G is not precisely zero. Especially it is possible to overcompensate $B_0$, which leads to a paramagnetic effect also in the ZFC sequence (Fig.\,\ref{Abb:PME}b).

\end{document}